\documentclass[]{spie}  

\usepackage{amsmath,amsfonts,amssymb}
\usepackage{graphicx}
\usepackage[colorlinks=true, allcolors=blue]{hyperref}

\title{Full Stokes polarimetry using Dual-Frequency Liquid Crystals}

\author[a]{K. Nagaraju}
\author[a]{D. V. S. Phanindra}
\author[b]{S. Krishna Prasad}
\author[b]{D. S. Shankar Rao}
\author[a]{P. Sreekumar}
\affil[a]{Indian Institute of Astrophysics, Koramangala II Block, Bengaluru, India.}
\affil[b]{Center for Nano and Soft-Matter Sciences, Jalahalli, Bengaluru, India.}

\authorinfo{Send correspondence to nagarajuk@iiap.res.in\\
Author Information:
K. Nagaraju: E-mail: nagarajuk@iiap.res.in\\ D. V. S. Phanindra:  E-mail: phanindra@iiap.res.in\\ S. Krishna Prasad: skprasad@cens.res.in\\D. Shankar Rao: shankar@cens.res.in\\P. Sreekumar:  E-mail: sreekumar@iiap.res.in}

\begin{document} 
\maketitle

\begin{abstract}
In a dual-frequency liquid crystal (DFLC), when the frequency of the applied voltage is more than a critical value ($f_c$), the dielectric anisotropy of the material changes from positive to negative. 
This causes the director to switch its orientation from parallel to the field (for $f < f_c$), to perpendicular to it ($f > f_c$).  
Hence DFLC can be used in modulating the light by switching the frequency of an externally applied voltage. 
We present in this work about application of DFLCs in full Stokes polarimetery. A polarization modulator has been worked out based on two DFLCs and two static retarders. The combination of DFLCs' switching and static retarders are chosen such that more or less equal weightage is given to all the Stokes parameters. Initial results on the optimization of position angles of the modulators are presented towards the goal of achieving polychromatic modulator in the wavelength range 600-900 nm.
\end{abstract}


\section{INTRODUCTION}
\label{sec:intro}  

Accurate information on the magnetic field, simultaneously at different
heights in the solar atmosphere is required to understand many physical processes which
take place on the Sun such as  explosive events (flares and CMEs), chromospheric and coronal heating, solar wind acceleration and etc. 
Multi-line spectropolarimetry is a powerful observational tool in
remote sensing the magnetic field through Zeeman or Hanle diagnostics for this purpose. Magnetic field measurements at the photosphere have become more or less routine but a similar level of maturity is yet to be achieved with respect to the measurements of the chromospheric magnetic fields\cite{lagg15}. 
Some of the reasons are the polarization signal in the spectral lines formed in the
chromosphere is low (in the order of $10^{-3}$ or less) and interpreting the measurements is challenging as NLTE effects have to be taken into account while modeling the spectral lines\cite{delacruzrodriguez17}.

In ground based observations seeing induced spurious polarization is a major obstacle in achieving high precision in polarization measurements.
This also known as seeing induced cross-talk \cite{lites87} is caused when consecutive measurements are combined to get the Stokes parameters. 
A dual-beam setup, in which orthogonally polarized beams are recorded simultaneously on two different detectors or on two different pixels of the same
detector, is widely being used for reducing the seeing induced cross-talk. 
However, dual-beam polarimetry suffers from differential gain effects which
is limiting the polarimetric sensitivity to $\approx 10^{-3}$ \cite{snik14}. 
Moreover, dual-beam setup only helps in reducing the cross-talk from Stokes-$I$ to 
Stokes-$Q$, $U$ and $V$ and the cross-talk among Stokes~$Q$, $U$ and $V$ still remains.
Another way of reducing seeing induced cross-talk is to carry out the measurements with fast modulation. 
Increase in modulation frequency systematically reduces seeing
induced cross-talk \cite{lites87,krishnappa12}. 

Various modulators are being used  such as Piezo-elastic modulators, Liquid Crystal Variable Retarders (LCVRs), Ferro-electric Liquid Crystals (FLCs) as well as rotating waveplate towards achieving fast modulation. Each of these modulators have their own pros and cons (see for e.g. \citenum{snik14}). \citenum{golovin03} and \citenum{Nie06} have shown that the  Dual-Frequency Liquid Crystals (DFLCs)  can be used as electrically switchable retarders by switching the LC drive frequency.
We, in this paper explore DFLCs as potential fast modulators in the context of full Stokes polarimetry. 
Some of the properties of DFLCs' that motivated us to explore them as modulators are: 1. They switch faster than conventional LCVRs, 2. Unlike FLCs their apertures
can be quite large. FLC sizes are limited because they are vulnerable to
mechanical stresses which cause defects across the aperture of FLC.

\section{Some of the relevent properties of DFLCs as polarization modulators}
\begin{figure} [ht]
   \begin{center}
   \includegraphics[height=16cm,angle=-90]{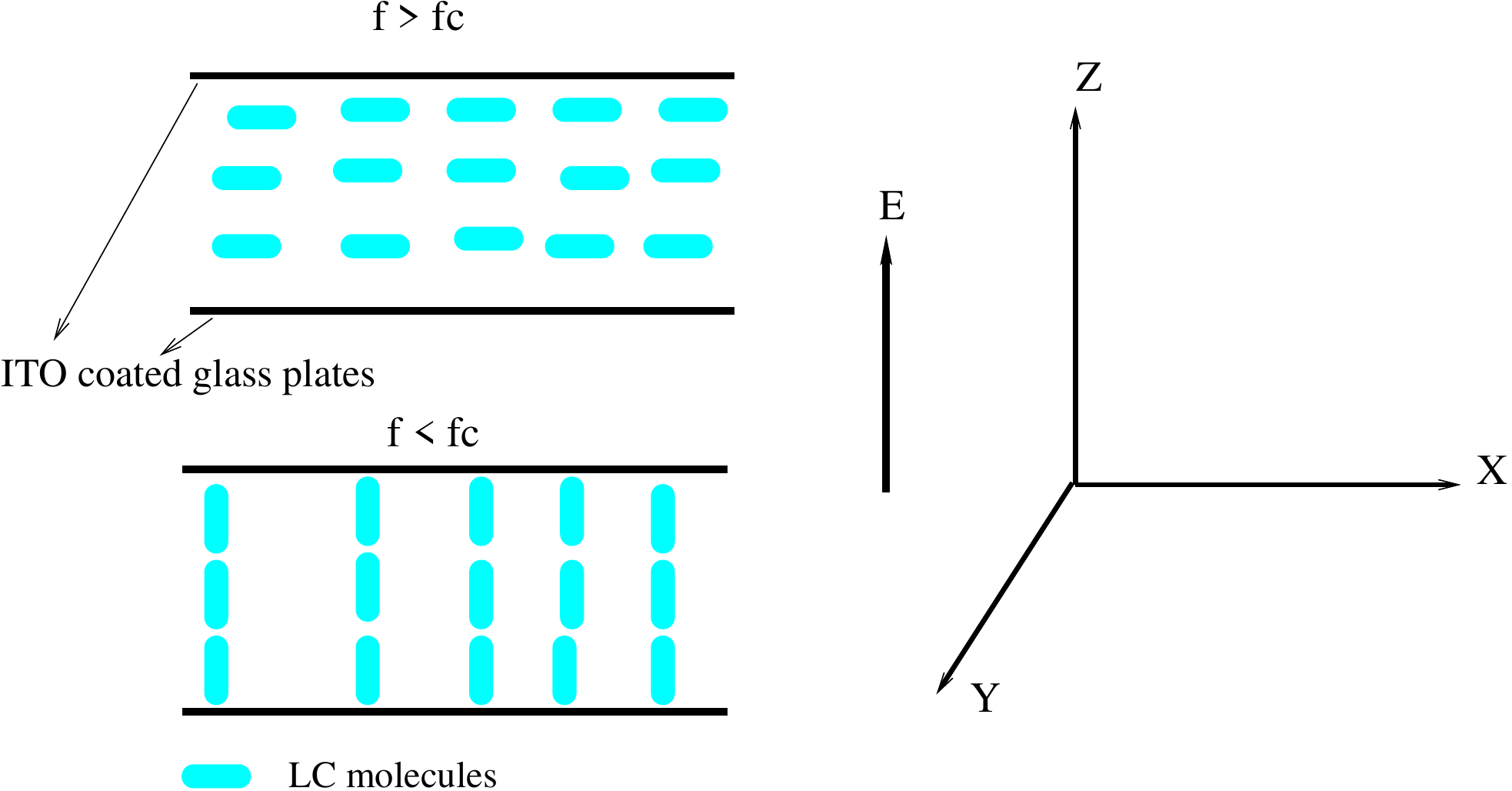}
   \end{center}
   \caption[example] 
   { \label{fig:LCcell} 
   A block diagram showing LC crystal cells in which the LC molecules are arranged perpendicular to the externally applied electric field $\boldsymbol{E}$ when $f>f_c$ and parallel to it when $f<f_c$. For reference a cartesian coordinate system is shown on the right side of the figure. The externally applied voltage is in the Z-direction, which is same as the light propagation direction.}
   \end{figure}

Dual-frequency liquid crystals are a type of nematic liquid crystals in which the dielectric
anisotropy changes from positive to negative when the frequency of externally applied voltage
pulses changes from $f<f_c$ to $f>f_c$\cite{schadt82}. The dielectric anisotropy is defined as the difference
in permittivity ($\epsilon$) in parallel and perpendicular directions to  the crystal
director {$\boldsymbol n$}, i.e. ($\Delta\epsilon=\epsilon_{\parallel}-\epsilon_{\bot}$).
The cross-over frequency '$f_c$' is the drive frequency of an externally applied voltage at which $\Delta\epsilon=0$. 
The crystal director '$\boldsymbol{n}$' aligns itself in the direction of the externally applied
electric field when $\Delta\epsilon$ is positive (when $f<f_c$) and perpendicular to it when $\Delta\epsilon$ is negative (when $f>f_c$). This process of switching is demonstrated in a block diagram shown in Fig.~\ref{fig:LCcell}. For reference, a coordinate system is shown on the
right side of the figure. The schematic shows that the LC material is sandwiched between
two ITO coated glass plates and the rubbing
on the glass plates is assumed to be in the X-direction so that the long axes of the LC
molecules are in the X-direction. Here we don't assume any pre-tilt angle which is normally used for increasing the switching speed. The light propagation direction is assumed to be
in the Z-direction which is same as the direction of externally applied electric field.
The LC configuration in the top panel shows the condition in which the externally applied
voltage has a frequency $f>f_c$ and the long axes of the molecules are arranged
perpendicular to the externally applied electric field '$\boldsymbol{E}$'. 
The bottom cell shows the condition in which $f<f_c$ and the long axes of molecules are arranged parallel to ${\boldsymbol E}$.

For the crystal configuration shown in Fig.~\ref{fig:LCcell} and the light propagating in the Z-direction, the crystal acts like a waveplate when $f>f_c$ and as a simple transparent glass plate when $f<f_c$. A finite phase delay between the X- and Y-components of the electric vector of the light beam is introduced in the former condition and no phase delay in the later condition. 
When the crystal molecules are aligned in the direction of light propagation direction, light sees azimuthally symmetric molecular structure hence there is no phase delay and the LC cell acts like a simple transparent medium. The case in which LC molecules are aligned perpendicular to the beam direction, there is a finite phase delay introduced between the X- and Y- components of the electric vector of the light beam as the molecules are aligned in the X-direction which causes azimuthal asymmetry. 
This property makes DFLC acts like a retarder when $f>f_c$.
We would like to note here that in practical terms we will not achieve zero retardance when
$f<f_c$ because not all the molecules will align in the direction of electric field. Because of strong surface anchoring the molecules close to the surface of the glass plates may not switch their direction. Apart from this LC molecules mayn't be fully azimuthally symmetric. However, in the following we assume ideal behaviour of DFLCs to design a modulator.

\section{Polarization modulator based on DFLCs}
 As we have seen in the previous section that the DFLCs offer finite retardance in the off state ($f>f_c$) and no phase delay in the on state ($f<f_c$). 

\begin{figure} [ht]
	\begin{center}
		\includegraphics[height=17cm,angle=-90]{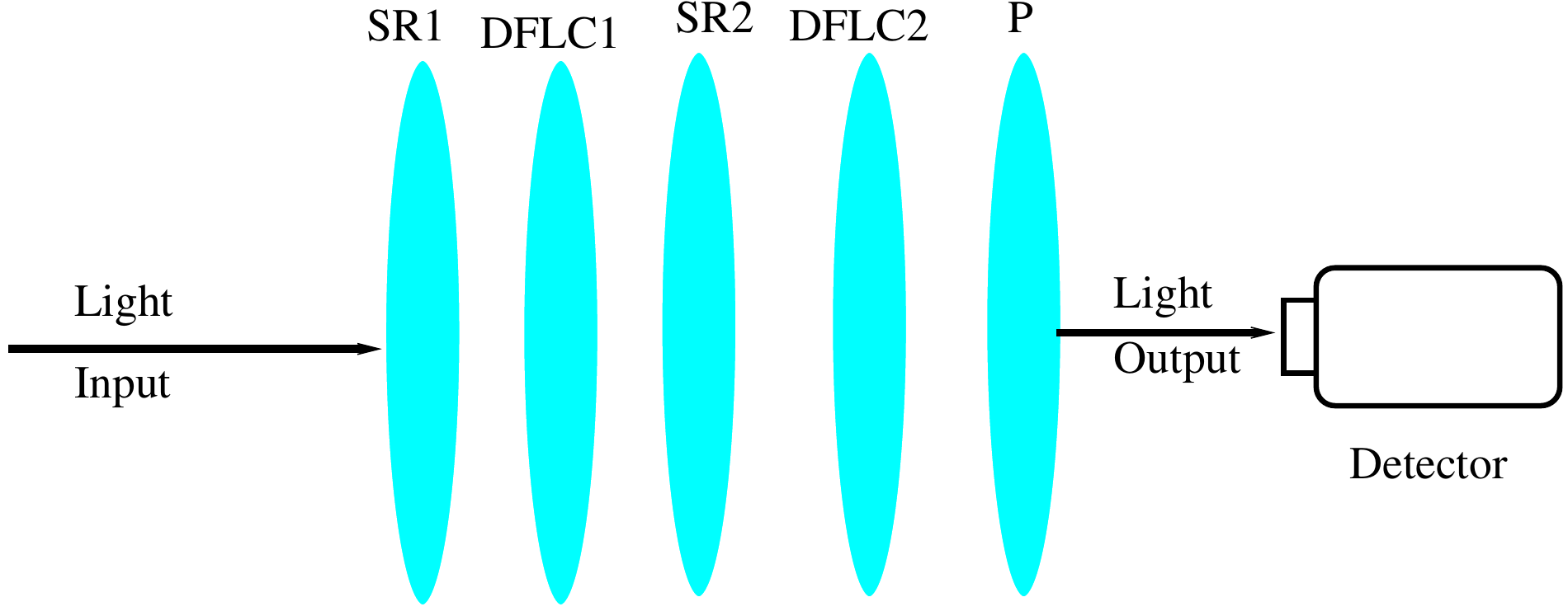}
	\end{center}
	\caption[example] 
	{ \label{fig:modsch} 
		Schematic of a modulator based on two DFLCs and two static retarders followed by a 
		polarization analyzer. SR1 and SR2 are the static retarders and P represent the polarization analyzer.}
\end{figure}

For full Stokes polarimetry we require a minimum of two DFLCs, as in combination they provide four distinct states of modulation which enables us to obtain four independent intensity measurements.
Minimum of four intensity measurements are required to derive four Stokes parameters.
Since, DFLCs offer no retardance in their on states we are required to
use additional waveplates in order to workout a balanced modulation scheme with a minimum number of modulation steps.
The modulator we propose is based on two DFLCs interlaced between two static retarders.
The static in the current context means the retardance of a given retarder is fixed. 
A schematic of the polarization modulator is shown  in Fig.~\ref{fig:modsch}. 
In this setup we have chosen DFLC1 and DFLC2 as half-waveplates and SR1 and SR2 as quarter-waveplates. The position angles of SR1, DFLC1, SR2 and DFLC2 are 45, 45, 22.5 and 67.5 DEGREES, respectively. The position of DFLC1 and SR1 in the beam can be interchanged which doesn't make any difference to the modulation scheme. On the other hand the positions of the DFLC2 and SR2 can not be interchanged.

The modulation scheme for the modulator setup (shown in Fig.~\ref{fig:modsch}) is given in Table~\ref{tab:modsch}. In this scheme Stokes-$U$ and $V$ have same weightage and Stokes-$Q$ has slightly higher weightage. If we choose SR2 as half-waveplate instead of a quarter-waveplate then weights of $Q$ and $V$ will be interchanged.

\begin{table}[ht]
\begin{center}       
	\begin{tabular}{|l|l|l|} 
		\hline
		\rule[-1ex]{0pt}{3.5ex}  DFLC1  & DFLC2 & Modulated Intensity  \\
		\hline
		\rule[-1ex]{0pt}{3.5ex}$f1<f_c$, $\delta=0$ & $f2>f_c$,$\delta=\pi$ & $I_1=I-0.7Q-0.5U+0.5V$ \\
		\hline
		\rule[-1ex]{0pt}{3.5ex}$f1>f_c$, $\delta=\pi$ & $f2>f_c$,$\delta=\pi$ & $I_2=I+0.7Q-0.5U-0.5V$ \\
		\hline
		\rule[-1ex]{0pt}{3.5ex}$f1>f_c$, $\delta=\pi$ & $f2<f_c$,$\delta=0$ & $I_1=I+0.7Q+0.5U+0.5V$   \\
		\hline
		\rule[-1ex]{0pt}{3.5ex}$f1<f_c$, $\delta=0$ & $f2<f_c$,$\delta=0$ & $I_1=I-0.7Q+0.5U-0.5V$  \\
		\hline 
	\end{tabular}
	\caption{Table showing a modulation scheme based on switching the drive frequencies of the DFLCs. In the above setup we have chosen DFLC1 and DFLC2 as half-waveplates and SR1 and SR2 as quarter-waveplates. The position angles of SR1, DFLC1, SR2 and DFLC2 are 45, 45, 22.5 and 67.5 DEGREES, respectively. $f1$ and $f2$ are the drive frequencies and $f_c$ is a criticial frequency for a given DFLC. If we choose SR2 as half-waveplate instead of a quarter-waveplate then efficiencies of $Q$ and $V$ will be interchanged.} 
	\label{tab:modsch}
\end{center}

\end{table}

\subsection{Polarimetric efficiency}

In Mueller matrix formalism, any polarization optics can be expressed as a $4\times4$ matrix. The modified Stokes vector can be written as
\begin{equation}
S_{out} = {\boldsymbol M} S_{in},
\end{equation}
where, $S=[I,Q,U,V]^T$ is a Stokes vector with the subscript $_{in}$ and $_{out}$ representing the input and output beams and $\boldsymbol{M}$ is the Mueller matrix of the optical component. The optical detectors are sensitive only to the first element
of $S_{out}$. Hence only the first row of the Mueller matrix $\boldsymbol{M}$ is relevant.
In order to measure four Stokes parameters we need a minimum of four intensity measurements with different configurations of the modulator. The modulated intensities are related to the input
Stokes vector through \cite{delToroIniesta03}
\begin{equation}
I' = {\boldsymbol O} S_{in},
\end{equation}
where $I'=[I_1,I_2...I_m]^T$, with $^T$ representing transpose operation, are the modulated intensities corresponding to $m$ modulation steps. ${\boldsymbol O}$ is a $m\times4$ array  called as modulation matrix.  The optimal demodulation matrix is the Moore-Penrose  pseudo-inverse of 
the modulation matrix ${\boldsymbol O}$\cite{delToroIniesta00},
\begin{eqnarray}
S_{in} &=& {\boldsymbol D} ~I':\\\nonumber
{\boldsymbol D} &=& \left({\boldsymbol O}^T {\boldsymbol O}\right)^{-1}{\boldsymbol O}^T.
\end{eqnarray}
The polarization efficiencies are defined as
\begin{equation}
\epsilon_k = \left(n \sum_{l=1}^n {\boldsymbol D}^2_{kl}\right)^{-1/2}.
\end{equation}
The efficiencies are such that 
\begin{equation}
\epsilon_1\le 1, ~\sum_{k=2}^4 \epsilon_k^2 \le 1.
\end{equation}

The modulation matrix corresponding to the modulation scheme described in Table~\ref{tab:modsch} is given by
\begin{equation}
O=
\begin{pmatrix}
1 & -0.7 & -0.5 & 0.5 \\
1 & 0.7 & -0.5 & -0.5 \\
1 & 0.7 & +0.5 & 0.5 \\
1 & -0.7 & +0.5 & -0.5   \\
\end{pmatrix}.
\end{equation}

The corresponding efficiencies are 
$(\epsilon_{Q},\epsilon_U,\epsilon_V)=(0.7,0.5,0.5)$.
The efficiencies of $Q$ and $V$ can be interchanged by choosing SR2 as a half-waveplate
instead of a quarter-waveplate. 

\section{Polychromatic modulator}
\begin{figure} [ht]
	\begin{center}
		\includegraphics[height=16cm,angle=-90]{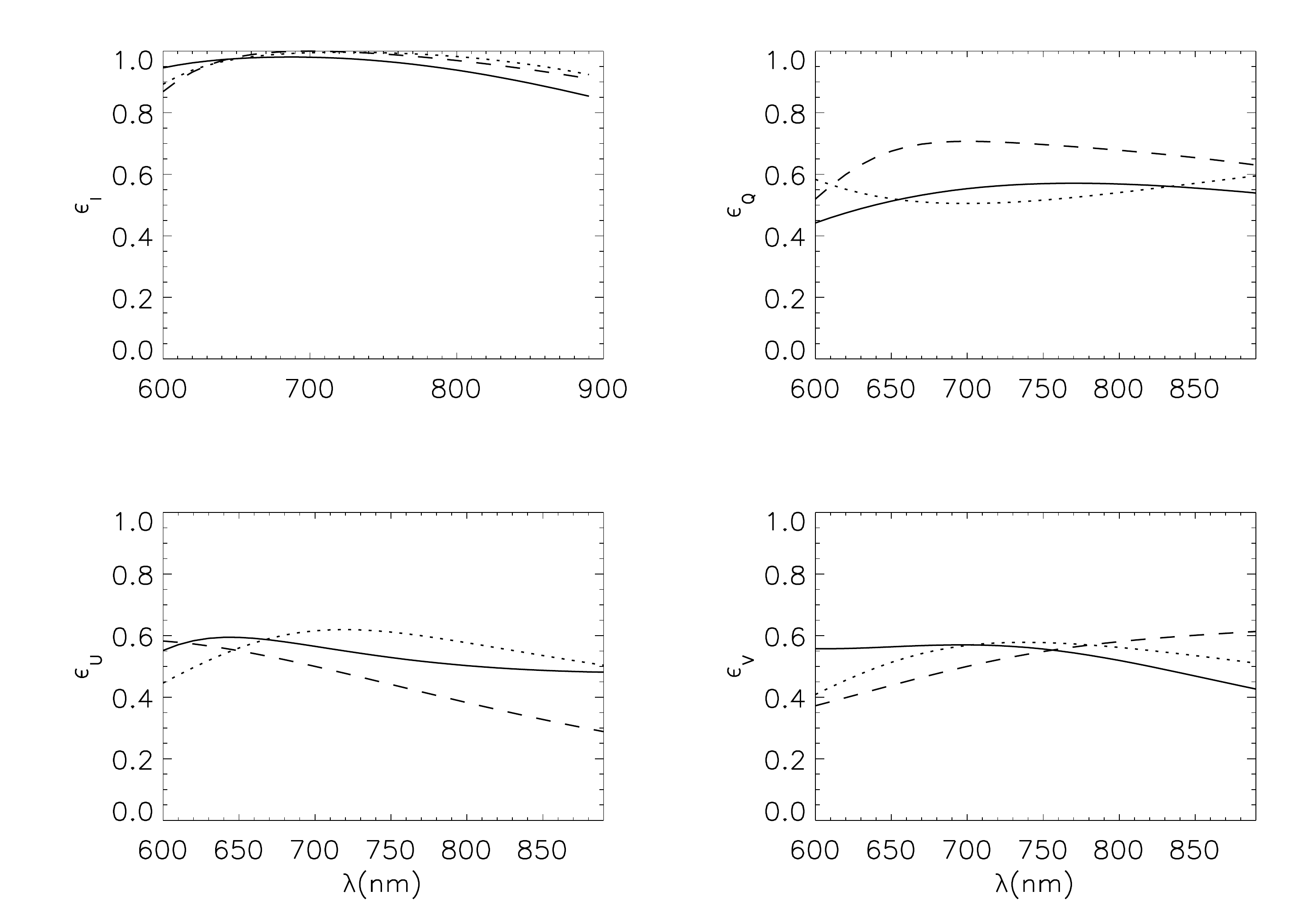}
	\end{center}
	\caption[example] 
	{ \label{fig:plotseff} 
	Plots of polarimetric efficiencies of Stokes parameters as a function of wavelength in the range from 600-900~nm. The design wavelength is 700~nm. The dotted lines correspond to the efficiencies without optimisation. The solid curves correspond to the efficiencies with both position angles and the retardances of the modulator optics are optimized. The dotted curves correspond to the case for which only the position angles optimized.}
\end{figure}

As we have seen in the previous section that the modulator efficiencies of Stokes-$Q$, $U$ and $V$ are not fully balanced. Apart from this, because of the chromatic nature of the modulator optics the polarimetric efficiencies are wavelength dependent. Plots of the efficiencies of Stokes parameters as a function of wavelength are shown in Fig.~\ref{fig:plotseff} for the modulator
setup described in Fig.~\ref{fig:modsch} and Table~\ref{tab:modsch}. The design wavelength is at 700~nm. The dashed curves correspond to the efficiencies for the angles and the retardances of the modulator optics as given in Table~\ref{tab:modsch} at the design wavelength. 
The retardances at other wavelengths are calculated under the assumption that
the retardance is inversely proportional to the wavelength (i.e. $\delta \propto 1/\lambda)$.

It has been shown in the past for other modulators based on for e.g. FLCs \cite{gisler03,tomczyk10}, static retarders \cite{snik14} and others that the efficiencies can be
achromatized by appropriately choosing the position angles and/or the retardances of the modulator optics. Finding optimum position angles and/or retardances is an optimization process. In this work we have also explored whether such a possibility exists for the
modulator discussed in this paper or not. As a first step towards this optimization we tried to find optimum
position angles which provide more or less uniform efficiencies in the wavelength range
600-900~nm. For finding the optimum position angles we adopted a direct search method. Which
means we systematically searched for the combination of position angles of the four modulator
optics in the range from $0-180^o$ in steps of $5^o$ using a computer program written in IDL. While searching we have used the constraints provided by two merit functions as defined below \cite{snik12}.
\begin{equation}
\label{eq:merita}
\Delta\epsilon_a =\left<
\sqrt{\left( \frac{1}{\sqrt{3}}-\epsilon_Q(\lambda) \right)^2 + \left(\frac{1}{\sqrt{3}}-\epsilon_Q(\lambda) \right)^2 + \left(\frac{1}{\sqrt{3}}-\epsilon_Q(\lambda) \right)^2 } \right>_{\lambda},
\end{equation}

where $\left<...\right>_\lambda$ denotes spectral averaging and
\begin{equation}
\label{eq:meritb}
\epsilon_b = \mathrm{max}\left(1-\sqrt(3) \mathrm{min}(\epsilon_Q(\lambda),\epsilon_U(\lambda),\epsilon_V(\lambda))\right).
\end{equation}

We have used the constraints that $\Delta\epsilon_a\le 0.1$ implying the variation in the efficiency of any of Stokes parameters does not exceed $10\%$ within the wavelength range considered and $\Delta\epsilon_a\le 0.3$ implying that the efficiency of each of the Stokes parameters is always $\ge 40\%$.
In a similar way we searched simultaneously for optimum position angles and retardances using the above constraints.
Plots of efficiencies as a function of wavelength when only the position angles are optimized as well as when both position angles and retardances are optimized are shown in Fig.~\ref{fig:plotseff}. 
The dotted curves correspond to the case for which only the angles are optimised and the solid curves correspond to the case for which both position angles-retardances are optimised. As we can see in these plots with optimization of angles and retardances the efficiencies approach achromaticity. This is quantified in the Table.~\ref{tab:meritfunct}.
The variation in the efficiencies is less than 10\% and the efficiencies are always above 40\% when the position angles and retardances are optimized, where as the variation in efficiency is as large as 20\% and the efficiency drops below 30\% when there is no optimization.
The case in which both position angles and retardances are optimized is not very different from when only the position angles are optmized. However, when we consider wider wavelength range we may see substantial difference. 

 \begin{table}[ht]
	\begin{center}       
		\begin{tabular}{|l|l|l|} 
			\hline
			\rule[-1ex]{0pt}{3.5ex} Cases & $\Delta\epsilon_a$ & $\Delta\epsilon_b$  \\
			\hline
			\rule[-1ex]{0pt}{3.5ex} 1 & 0.21 & 0.5  \\
			\hline
			\rule[-1ex]{0pt}{3.5ex} 2 &0.08 & 0.29   \\
			\hline
			\rule[-1ex]{0pt}{3.5ex}  3 &0.09 & 0.26 \\
			\hline
		\end{tabular}
		\caption{Values of merit functions as described in Eqs.~\ref{eq:merita}-\ref{eq:meritb} for the cases when no optimization (Case 1), with angles only optimized (Case 2) and with both angles and retardances are optimized (Case 3).} 
		\label{tab:meritfunct}
	\end{center}
\end{table}

\section{Conclusions}
Keeping in view of the potential application of DFLCs in polarimetry as they switch faster than the conventional LCVRs and the apertures can be much larger than the FLCs, a full Stokes polarization modulator based on DFLCs has been worked out.  The proposed modulator has two DFLCs interlaced between two static retarders. The modulation scheme presented in this
paper is very close to a balanced scheme with 70\% efficiency for Stokes-$Q$ and 50\% efficiency for Stokes-$U$ and $V$.
While working out a modulation scheme we have assumed that the retardance of a DFLC varies between 0 (for $f<f_c$) and a finite value ($\delta$: for $f>f_c$). In practice, the molecules close to the surface may not switch completely due to strong anchoring to the substrate plates because of which the retardance may not be completely zero. While designing a modulator this aspect has to be taken into account.
 

A polychromatic design of the modulator based on DFLCs has been presented. It has been found that by optimizing the position angles and the retardances, it is possible to achieve a polychromatic modulator very similar to those based on FLCs \cite{gisler03,tomczyk10} or multiple waveplates \cite{snik12}. 
The optimization method, which is a direct search method, presented
in this paper is very time consuming. We are in the process of adopting a more efficient method for finding the optimum position angles and the retardances of modulator optics.
\bibliography{report} 
\bibliographystyle{spiebib} 

\end{document}